\documentstyle[aps,tighten,epsf]{revtex}

\begin{document}

\draft

\title{A semi--classical over--barrier model for charge exchange 
between highly charged ions and one--optical electron atoms }

\author{Fabio Sattin \thanks{E-mail: sattin@igi.pd.cnr.it}}

\address{Consorzio RFX, Corso Stati Uniti 4, 35127 Padova, ITALY}

\maketitle
\abstract{
Absolute total cross sections for electron capture between slow, highly 
charged ions and alkali targets have been recently measured. It is found 
that these cross sections follow a scaling law with the projectile charge 
which is different from the one previously proposed basing on a  
classical over--barrier model (OBM) and verified using rare gases and 
molecules as targets. In this paper we develop a "semi--classical" 
(i.e. including some quantal features) OBM attempting to recover experimental 
results. The method is then applied to ion--hydrogen collisions and 
compared with the result of a sophisticated quantum-mechanical calculation. 
In the former case the accordance is very good, while in the latter one
no so satisfactory results are found.
A qualitative explanation for the discrepancies 
is attempted.}
 

\section{Introduction}
The electron capture processes in collisions of slow, highly charged 
ions with neutral atoms and molecules are of great importance not only 
in basic atomic physics but also in applied fields such as fusion 
plasmas and astrophysics. \\
In the past years a number of measurements have 
been carried on the collisions between highly charged ions and rare 
gases \cite{uno} or molecules \cite{due}, in which one or several electrons were 
transferred from the neutral target to a  charged projectile:
\begin{equation}
\label{eq:collision1}
A^{+q} + B  \to A^{(q-j)+} + B^{j+} \quad .
\end{equation}
 Their results-together with those from a number of other laboratories-yielded 
a curve which can be fitted within a single scaling law
(a linear relationship) when plotting cross section 
$\sigma$ {\it versus } projectile charge  $q$: 
it is almost independent of the projectile species and of the impact velocity $v$ 
(at least in the low--speed range $ v < 1$ au). When one extends 
experiments to different target species, the same linear relation holds between 
$\sigma$ and $q/I_t^2$, with $I_t$ the ionization potential of 
the target \cite{tre,quattro}. \\
It is found that this scaling law could to be predicted,  in the limit
of very high projectile charge,  by a modification of an extended 
classical over--barrier model (ECBM), allowing for multiple electron capture, 
proposed by Niehaus \cite{cinque}. Quite recently a confirmation of this 
scaling has come from a sophisticated quantum--mechanical calculation 
\cite{sei}.
 
\par
Similar experiments were carried on more recently for collisions between ions
and alkali atoms \cite{sette}. The results show that the linear trend is roughly 
satisfied, but the slope of the straight line is grossly overestimated by the ECBM:
in Fig. 1 we show some data points (stars with error bars) together with 
the analytical curve from the ECBM (dashed curve) which, for 
one--electron atoms, is written \cite{tre,quattro}
\begin{equation}
\label{eq:niehaus}
\sigma = 2.6\times 10^{3} q/I_{t}^{2}   \quad [10^{-20}{\rm m}^{2}] 
\end{equation}
($I_{t}$ in eV). 
It should be noticed that experimental data are instead well fitted 
by the results of a Classical Trajectory MonteCarlo (CTMC) code \cite{sette}.

\par
The ECBM of ref. \cite{tre} works
in a simplified one-dimensional geometry where the only
physically meaningful spatial dimension is along the internuclear axis.
It does not take into account the fact that the electrons 
move in a three-dimensional space. This means that only 
a fraction of the electrons actually can fulfil the conditions 
dictated by the model. For rare gases and molecules, which have a 
large number of active electrons, this can be not a trouble 
(i.e., there are nearly always one or more electrons which can participate 
to the collision). For alkali atoms with only one
active electron, on the other hand, an overestimate of the capture
probability by OBM's could be foreseen.

\par
With present--days supercomputers there are relatively few difficulties in 
computing cross sections from numerical integration of the time-dependent
Schr\"odinger equation (e.g. refer to ref. \cite{sei}). Notwithstanding this, 
simple models are still valuable since they allow to get analytical 
estimates which are easy to adapt to particular cases, and give 
physical insight on the features of the problem. For this reason new 
models are being still developed \cite{otto,nove}.\\
In this paper we present a modified OBM which allows 
to get a better agreement with the experimental data of ref. \cite{sette}.

\section{The model}
We start from the same approach as Ostrovsky \cite{otto} (see also 
\cite{ryufuku}): be {\bf r} the electron 
vector relative to the neutral atom ({\bf T}) and {\bf R} the internuclear 
vector between {\bf T} and the projectile {\bf P} 
(see Fig. 2 for a picture of the geometry: it is an adaptation of Figure 1 
from ref. \cite{otto}).
Let us consider the plane containing the electron, {\bf P} and {\bf 
T}, and use cylindrical polar coordinates $(\rho, z, \phi)$ to 
describe the position of the electron within this plane. We can choose
the angle $\phi = 0$ and the $z$ direction along the internuclear axis. 
We will assume that the target atom can be 
described as an hydrogenlike atom, which is not a bad approximation 
when dealing with alkali atoms. \\
The total energy of the electron is
\begin{equation}
\label{eq:uno}
E = {p^{2} \over 2 } + U = {p^{2} \over 2 }
-{ Z_{t} \over \sqrt{\rho^{2}+z^{2} } }- {Z_{p} \over \sqrt{\rho^{2}+(R-z)^{2}}} 
\quad .
\end{equation}
$Z_{p}$ and $Z_{t}$ are the charge of the projectile and the 
effective charge of the target seen by the electron, respectively, and 
we are using atomic units.\\
We can also approximate $E$ as 
\begin{equation}
\label{eq:due}
E(R) = - E_{n} - {Z_{p} \over R} \quad .
\end{equation}
$E_{n}$ is given by the quantum--mechanical value: $ E_{n} = 
Z_{t}^{2}/(2 n^{2})$. This expression is asimptotically correct
as $ R \to \infty$.
 
On the plane (e, {\bf P}, {\bf T}) we can draw a 
section of the equipotential surface 
\begin{equation}
\label{eq:equip}
 U(z,\rho,R) = - E_n  - {Z_p \over R} \quad . 
\end{equation}
This represents the limit of the region classically allowed to the electron.
When $R \to \infty$ this region is decomposed into two 
disconnected circles centered around each of the two nuclei. Initial
conditions determine which of the two regions actually the electron lives 
in. \\
As $R$ diminishes there can be eventually a time where the two regions
become connected.  It is easy to solve eq. (\ref{eq:equip}) for $R$
by imposing that $\rho_{m} = 0$ and that there must be an unique solution 
for $z$ with $ 0 < z < R$:
\begin{equation}
\label{eq:rm}
R_{m} = { Z_{t} + 2 \sqrt{Z_{t} Z_{p}} \over E_{n} } \quad .
\end{equation} 
In the spirit of OBMs it is the opening of 
the equipotential curve between {\bf P} and {\bf T}  
which leads to a leakage of electrons from one nucleus to another, and 
therefore to charge exchange. 
Along the internuclear axis the potential $U$ has a maximum at
\begin{equation}
z = z_{0} = R { \sqrt{Z_{t}} \over \sqrt{Z_{p}} + \sqrt{Z_{t}} } 
\quad .
\end{equation} 
Whether the electron crosses this potential barrier 
depends upon its initial conditions. These are chosen from a 
statistical ensemble, which we will leave unspecified for the moment.  
Let $N_{\Omega}$ be the 
fraction of trajectories which lead to electron loss at the time $t$ 
and $W(t)$ the probability for the electron to be still bound to the 
target, always at time $t$. The fraction of losses in the interval 
$t, t + dt$ is given by 
\begin{equation}
\label{eq:losses}
dW(t) = - N_{\Omega} {dt \over T_{em}} W(t) \quad , 
\end{equation}
with $T_{em}$ the period of the electron motion along its orbit. 
A simple integration yields the leakage probability
\begin{equation}
\label{eq:prob}
P_l = 1 - \exp \left( - {1 \over T_{em}} \int_{-\infty}^{+\infty} N_{\Omega} dt
\right) \quad .
\end{equation}
In order to actually integrate Eq. (\ref{eq:prob}) we need to know the 
collision trajectory; an unperturbed straight line with $b$ impact parameter 
is assumed:
\begin{equation}
\label{eq:traiettoria}
R = \sqrt{b^{2} + (v t)^{2}} \quad .
\end{equation}
At this point it is necessary to give an explicit expression for 
$N_{\Omega}$.  The electron is supposed to be in the ground state 
($n = 1, l = m = 0$). $T_{em}$ becomes therefore \cite{landau}
\begin{equation}
\label{eq:tempo} 
 T_{em}  = 2 \pi /Z_{t}^{3} \quad .
\end{equation}   
Ref. \cite{otto} adopts a geometrical reasoning: the classical electron trajectories, 
with zero angular momentum, are ellipses squeezed onto the target nucleus. 
The only trajectories which are allowed to escape are those whose aphelia 
are directed towards the opening within the angle $\pm \theta_{m}$.  The 
integration over this angle yields an analytical expression for 
$N_{\Omega}$ (Eq. 17 of ref. \cite{otto}).
In Fig. 1 we show the results obtained using Ostrovsky's model ( 
dotted curve--eqns. 8,17 of ref. \cite{otto})
\footnote{Beware of a small difference in 
notation between the present paper and \cite{otto}: here we use an 
effective charge for the target, $ Z_{t} = \sqrt{2 E_n}$, while
\cite{otto} uses an {\it effective quantum number} 
$ n_{t} = 1/\sqrt{2 E_n}$ with the effective charge of the target set to 1.}. 
Notice that from direct inspection of the 
analytical formula, one sees that the scaling law is not exactly 
satisfied, at least at small values of the parameter $q/I_{t}^{2}$, 
and this is clearly visible in the plot. The result
is almost equal to the scaling (\ref{eq:niehaus}). 
\par
The present approach is based on the electron {\em position} instead 
than on electron {\em direction }.
The recipe used here is (I) to neglect the dependence from the 
angle: all electrons have the same probability of escaping, regardless of 
their initial phase. Instead, (II) the lost electrons are precisely those 
which, when picked up from the statistical ensemble, are found farther 
from nucleus {\bf T} than the distance $z_{0}$:
\begin{equation}
\label{eq:integrale}
N_{\Omega} = \int_{z_{0}}^{\infty} f(r) dr \quad ,
\end{equation}
with $f(r)$ the electron distribution function. 

\par
There is not a unique choice for $f(r)$: the (phase-space) microcanonical 
distribution 
\begin{equation}
\label{eq:microcan}
\tilde{f}({\bf r},{\bf p}) \propto \delta\left( E_{n} + {p^{2} \over 2} - {Z_{t} 
\over r}  \right)
\end{equation}
($\delta$ is the Dirac delta) has been often used in literature since the 
works \cite{percival} as it is known that, when integrated over spatial coordinates, it 
reproduces the correct quantum--mechanical momentum distribution function 
for the case of the electron in the ground state \cite{dieci} (more recently 
the same formalism has been extended to Rydberg atoms  \cite{undici}).
After integration over momentum variables one gets instead
a spatial distribution function \cite{cohen}
\begin{equation}
\label{eq:classica}
f_{mc}(r) = 
 { Z_{t} (2 Z_{t})^{3/2} \over \pi } r^{2} \sqrt{ { 1 \over r} - 
{ Z_{t} \over 2} }   , \quad r < 2/Z_{t} 
\end{equation}
and zero elsewhere (The lowerscript "mc" is to emphasize that 
it is obtained from the microcanonical distribution).
However, this choice was found to give poor results.  
It could be expected on the basis of the 
fact that (\ref{eq:classica}) does not extend beyond $r = 2/Z_{t}$ 
and misses therefore all the large impact--parameter collisions.
In the spirit of the present approach, it should be instead important to have an 
accurate representation of the spatial distribution. We use therefore
for  $f(r)$ the quantum mechanical formula for an electron in the ground state:
\begin{equation}
\label{eq:funzione}
f_{1s}(r) = 4 Z_{t}^{3}  r^{2} \exp \left(- 2 Z_{t} r \right) 
\end{equation}
which, when substituted in (\ref{eq:integrale}), gives
\begin{equation}
\label{eq:frazione}
N_{\Omega} = \left[ 1 + 2 z_{0} Z_{t} + 2 (z_{0} Z_{t})^2 \right] 
\exp\left( -2 z_{0} Z_{t} \right) \quad .
\end{equation}
Since the choice for $f(r)$ does not derive from any classical 
consideration, we call this method a ``semi--classical'' OBM. \\
Notice that, in principle, one could go further and compute $f(r)$
from a very accurate wavefunction, fruit of quantum mechanical computations 
(see \cite{tredici}), but this is beyond the purpose of the present paper
(it could be worthy mentioning a number of other attempts of building
stationary distributions $f(r)$, mainly in connections with 
CTMC studies, see \cite{eich,hardie,montem}). \\
The $f(r)$ of Eq. (\ref{eq:funzione}) does not reproduce the correct 
momentum distribution, nor the correct energy distribution (which  
could be obtained only by using eq. (\ref{eq:microcan}). However, it 
is shown in \cite{cohen} that this choice gives an energy distribution  
for the electrons, $f(E)$, peaked around the correct value $E_{n}$, 
and $<E> = E_{n}$, where $<\ldots>$ is the average over $f(E)$.

\par
Some important remarks are to be done here. First of all, a 
question to be answered is: why use an unperturbed 
distribution, when the correct one should be sensitively modified by 
the approaching of the projectile. The answer is, obviously, that this
choice allows to perform calculations analitically. We are doing here a sort 
of classical counterpart of a quantum--mechanical Born calculation: there, too, 
the matrix elements are computed as scalar products over unperturbed states, 
regardless of any perturbation induced by the projectile.  
In the following, however, some considerations about possible 
improvements over this simple approximation will be done. \\
A second question regards the meaning of the factor $dt/T_{em}$ in
eq. (\ref{eq:losses}): in Ostrovsky's paper this is the fraction  of 
electrons which enter the loss zone during the time interval $dt$
and is valid under the hypothesis of a uniform distribution of 
initial phases of the electrons. In our case this this assumption 
ceases to be valid: electrons actually spend different fractions of 
their time at different radial distances from {\bf T}, depending on
their energy. We will do a (hopefully not too severe) assumption by assuming
that, on the average, the expression (\ref{eq:losses}) still holds.

\section{Results} 
\subsection{Iodine - Cesium}
This study has been prompted by the ion-atom experiments of 
\cite{sette}: first of all, therefore, we apply the above model to 
the process of electron capture
\begin{equation}
\label{eq:processo}
{\rm I}^{q+} + {\rm Cs} \to {\rm I}^{(q-1)+} + {\rm Cs}^{+}
\end{equation}
with $q = 6 \div 30$. Impact energy is $1.5\times q$ keV \cite{sette}.
The ionization potential of Cesium is $ I_{t} = 3.9 $ eV.
Solid line in Fig. 1 is the result of the present model: 
the agreement is fairly good.

\subsection{Bare ions - Hydrogen} 
As second test, we have computed cross section for captures
\begin{equation}
\label{eq:ho}
{\rm H} + {\rm O}^{8+} \to {\rm H}^{+} + {\rm O}^{7+}
\end{equation}
and
\begin{equation}
\label{eq:hhe}
{\rm H} + {\rm He}^{2+} \to {\rm H}^{+} + {\rm He}^{+}
\end{equation}
and compared it with similar calculations done using the molecular 
approach by Harel {\it et al} \cite{harel}. The results are summarized 
in fig. 3.  There is a sharp discrepancy in the behaviour for $v \to 
0$, where the present model predicts an increasing cross section.
At very low speed it is the concept itself of atomic distribution function 
which becomes questionable, and molecular aspects become important. 
Besides, quantum effects such as the discreteness of the energy levels
also play a major role and are completely missed by this approach. 
In the higher velocity part, the present model underestimates the more accurate 
value by a factor 2 for process (\ref{eq:ho}), but the error is much less, 
just 25 \%, for process (\ref{eq:hhe}). These two ions have been 
chosen {\it ad hoc}: they correspond to values of the ratio
$Z_{t}/Z_{p} = 1/8$ and 1/2 respectively.  
In the (I, Cs) test this ratio ranged from $\approx 1/12$ to 
$\approx 1/60$ depending upon the projectile charge. 
This means that in the former case the perturbation of the projectile 
on the electron distribution function is comparable to the (I, Cs) case, 
while in the latter it is much less. We expect the electron distribution
function to be more and more perturbed as $Z_{t}/Z_{p} \to 0$. \\

\section{Summary and conclusions}
We have developed in this paper a very simple OBM for charge exchange. 
It exploits some features of the quantum mechanical version of the 
problem, thus differing from similar models which are solely 
classical. The agreement with experiment is much better than previous 
calculations where a comparison could be made. It is far from excellent, 
but reasons for the (partial) failure have been suggested. \\
As it stands, the model is well suited for 
one-optical-electron atoms (since it uses hydrogen--like 
wavefunctions), therefore we do expect that other classical OBM's can 
still work better in the many-electrons targets studied in previous 
experiments. \\
Some improvements are likely to be added to the present model:
a possible line of investigation could be  coupling the present method 
with a very simplified calculation of the evolution of the wavefunction, using 
quantum mechanics. From this one should not compute the $f$ as coming 
from a  single state, but as a linear combination including also excited 
wavefunctions (the relative weights in the combination should be given 
by the quantum mechanical calculation). Work in this direction 
is currently underway.  

\section*{Acknowledgments}
It is a pleasure to thank the staff at National Institute for Fusion 
Science (Nagoya), and in particular Prof. H. Tawara and Dr. K. Hosaka 
for providing the data of ref. \cite{sette} and for useful discussions about 
the subject.
The referees through their suggestions and criticism have made 
the manuscript readable.

\newpage

\section*{Figure Captions}

\begin{figure}
\epsfxsize=12cm
\epsfbox{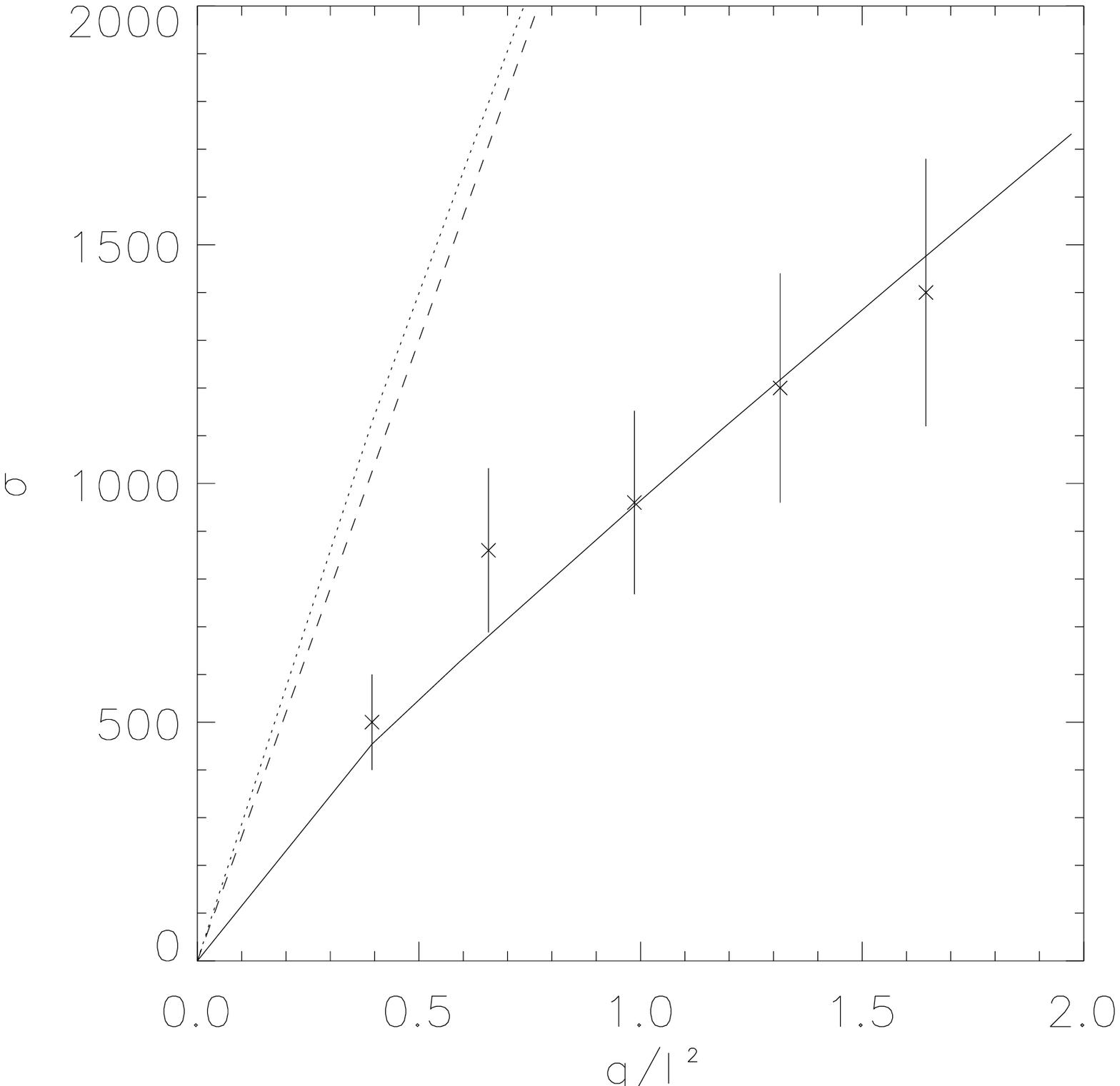}
\caption{Comparison between experimental data and prediction 
from models for electron capture cross section of process 
(\ref{eq:processo}). Stars, experiment with 20\% error bar; 
dashed line, scaling law from Niehaus (Eq. \ref{eq:niehaus});
dotted line, Ostrovsky's scaling law; 
solid line, scaling law from present model.
$\sigma$ is in $10^{-20}{\rm m}^{2}$, $I_{t}$ in eV.} 
\end{figure}

\begin{figure}
\epsfxsize=12cm
\epsfbox{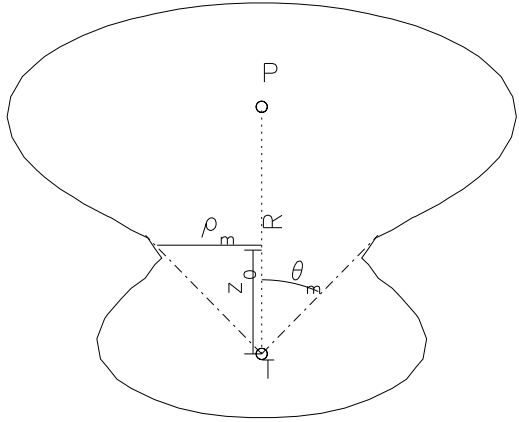}
\caption{Geometry of the scattering. {\bf P} and {\bf T} are the 
projectile and target nucleus respectively. The enveloping curve shows 
a section of the equipotential surface $U = E$, i.e., it is the border 
of the region classically accessible to the electron. $R$ is the 
internuclear distance. The parameter 
$\rho_{m}$ is the radius of the opening which joins the potential 
wells, $\theta_{m}$ the opening angle from {\bf T};
$z_{0}$ is the position of the potential's saddle point.}
\end{figure}

\begin{figure}
\epsfxsize=12cm
\epsfbox{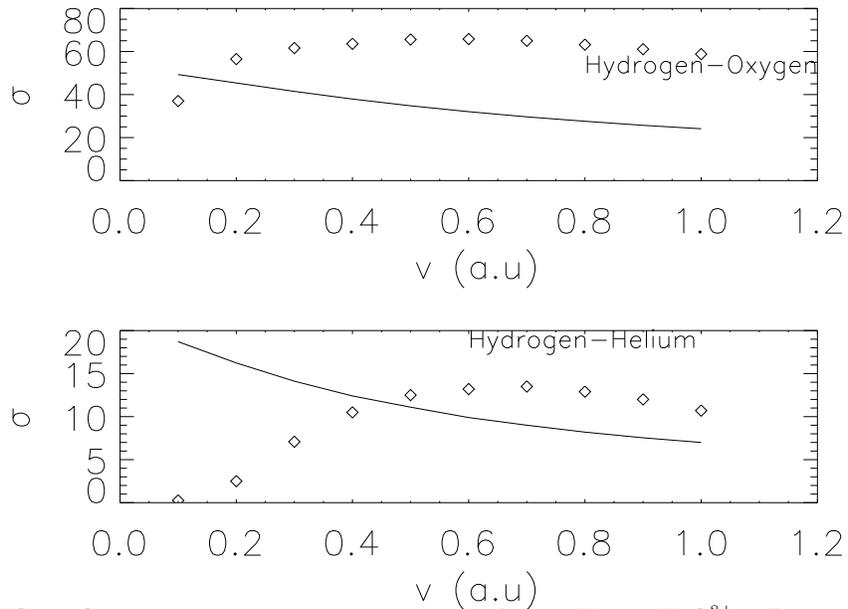}
\caption{Capture cross section {\it versus} impact velocity.
Upper, H--O${}^{8+}$ collisions; lower, H--He${}^{2+}$ collisions.
Diamonds, data from ref. 20; solid line, present model. }
\end{figure}

\end{document}